\documentclass[10pt,letterpaper]{article}

\usepackage{amssymb,amsmath}
\usepackage{overpic}
\usepackage{cite}
\usepackage{opex3}
\begin{document}

%% NOTE: TITLE PAGE & TOC NOT USED FOR MANUSCRIPT SUBMISSIONS %%

\title{Chiral metamaterials: retrieval of the effective parameters with and without substrate}

\author{Rongkuo Zhao$^{1,2}$,Thomas Koschny$^{1,3}$ and Costas M. Soukoulis$^{1,3,*}$}

\address{$^1$
Ames Laboratory and Department of Physics and Astronomy, Iowa State University, Ames, Iowa 50011, USA\\
$^2$
Applied Optics Beijing Area Major Laboratory, Department of Physics, Beijing Normal University, Beijing 100875, China\\
$^3$
Institute of Electronic Structure and Laser, FORTH, and Department of Materials Science and Technology, University of Crete, 71110 Heraklion, Crete, Greece
}
 \email{$^*$soukoulis@ameslab.gov}

%%% ABSTRACT

\begin{abstract}
After the prediction that strong enough optical activity may result in negative
refraction and negative reflection, more and more artificial chiral
metamaterials were designed and fabricated at difference frequency ranges from
microwaves to optical waves. Therefore, a simple and robust method to retrieve
the effective constitutive parameters for chiral metamaterials is urgently
needed. Here, we analyze the wave propagation in chiral metamaterials and
follow the regular retrieval procedure for ordinary metamaterials and apply it
in chiral metamaterial slabs. Then based on the transfer matrix technique, the
parameter retrieval is extended to treat samples with not only the substrate but also the top layers. After the
parameter retrieval procedure, we take two examples to check our method and study how the substrate influences on the thin chiral metamaterials slabs. We find that the substrate may cause the homogeneous slab to be inhomogeneous, i.e. the reflections in forward and backward directions are different. However, the chiral metamaterial where the resonance element is embedded far away from the substrate is insensitive to the substrate. 
\end{abstract}

 % REPLACE WITH CORRECT OCIS CODES FOR YOUR ARTICLE

%%%%%%%%%%%%%%%%%%%%%%%%%%%%%%%%%%%%%%%%%%%%%%%%%%%%%%%%%%%%%%%%%%%%%%%%%%%%%%%
%
\section{Introduction}
\label{introduction}
Chiral metamaterials (CMMs), which are artificial materials that lack any
planes of mirror symmetry, possess strong ability to rotate the plane of
polarization of electromagnetic waves. 
This ability is called \textit{optical activity} which is of great interest 
to many areas of science, for example, analytical chemistry and molecular biology. 
Recently, it was predicted that strong enough optical activity may result 
in negative refraction \cite{pendry, Tretyakov,Monzon,TretyakovPNFA,YannopapasJPCM,AgranovichPRB} and negative reflection
\cite{negativereflection}. After these predictions \cite{pendry, Tretyakov,Monzon,TretyakovPNFA,YannopapasJPCM,AgranovichPRB,negativereflection}, a lot of artificial CMMs were designed and fabricated at
different frequency ranges from microwaves to optical frequencies
\cite{chiral_Martin, chiral_Plum, chiral_Zhou, chiral_Dong,
chiral_Plum_APL2008, chiral_Plum_PRL2009, chiral_Jelinek, chiral_Bingnan,
chiral_Yannopapas, chiralexperiments_Zhang, chiralexperiments_Plum,
chiralexperiments_Gonokami, chiralexperiments_Decker_2010,chiralexperiments_Decker_2009,
chiralexperiments_Decker_2007}. The basic physical quantities to characterize
the optical properties of CMMs are the effective constitutive parameters.
Therefore, a simple and robust method to retrieve the effective parameters for
CMMs is urgently needed. 

Parameter retrieval \cite{retrievalref_Smith_1, retrievalref_Chen,
retrievalref_Koschny_1, retrievalref_Koschny_2, retrievalref_Smith_2,
retrievalref_Li} is a basic and important technique to obtain the
electromagnetic properties of the effective media. 
The effective media are usually considered to be homogeneous when the 
size of structure is much smaller than the wavelength $\lambda$. 
The retrieval parameters, $\epsilon$ and $\mu$ or $n$ and $Z$,  
are well defined and can be determined from reflection
and transmission coefficients ($S$ parameters)\cite{retrievalref_Smith_1,
retrievalref_Chen, retrievalref_Koschny_1, retrievalref_Koschny_2,
retrievalref_Smith_2, retrievalref_Li}. 
We follow the similar procedure\cite{retrievalref_Smith_1} to do the 
parameter retrieval for CMMs as it was done in ordinary metamaterials. The procedure of the retrieval for free-standing CMM slabs was briefly shown in our previous paper\cite{BingnanJOA}. However, in most of the experiments\cite{chiralexperiments_Zhang, chiralexperiments_Decker_2007,
chiralexperiments_Plum, chiralexperiments_Gonokami, chiralexperiments_Decker_2010, chiralexperiments_Decker_2009},
the CMMs are fabricated on a substrate.
Therefore, a retrieval procedure for samples with substrates should be developed. 
This is an important development for obtaining the electromagnetic
properties of CMMs from experimental data. 
Kwon et al.\cite{retrievalwithsubstrate} presented a retrieval procedure for 
CMMs with a semi-infinite substrate. 
However, experimentally the thickness of the substrate is finite. 
Here, we present a simple and robust retrieval method based on the 
transfer matrix technique. It can deal with samples not only with substrates but also with top layers. 

After the parameter retrieval, we give two chiral metamaterial examples: 1) twisted-crosses and 2) four-folded rotated $\Omega$-particles. 
For the twisted-crosses CMMs, the resonance elements are very close to the substrate. The existence of the substrate, therefore, has a huge influence on the thin CMM slab. It causes the resonance shifts to low frequencies and induces the homogeneous slab to be inhomogeneous. In order to correct the inhomogeneity, we need to introduce a top layer, the same material as the substrate, to cover the CMM slab. The top layer will red-shift the resonance further. For the four-folded rotated $\Omega$-particles CMMs, the resonance elements are embedded far away from the substrate. The influence of the substrate can be neglected. In the meanwhile, profiting from the retrieval method and the precise simulation, we fit the retrieval results using the analytical formulas deriving from the $\Omega$-particle resonator model. They agree with each other very well.  

The following sections are arranged as follows. 
In Sec. \ref{retrieval}, we first obtain explicit analytic expressions of the retrieval parameters
for free standing CMM slabs, then we extend this method to deal with the samples 
with the substrate and top layers. In Sec. \ref{examples}, we use the parameter retrieval methods to study two CMM examples and obtain the influence of substrate/top layer on thin layer CMM slabs. Finally in Sec. \ref{conclusion}, we present our conclusions.

%%%%%%%%%%%%%%%%%%%%%%%%%%%%%%%%%%%%%%%%%%%%%%%%%%%%%%%%%%%%%%%%%%%%%%%%%%%%%%%
%
\section{Retrieval method}
\label{retrieval}

% ----------------------------------------------------------------------------
%
\subsection{Eigen waves in chiral materials}
For reciprocal (Pasteur) chiral materials \cite{lindell, Serdyukov}, the
strength of the cross coupling between the magnetic and electric field is
characterized by the parameter $\kappa$ called \textit{chirality}. 
The constitutive relations in chiral medium are usually written 
as \cite{constitutive}:
\begin{equation}
\label{constitutive}
\left(
\begin{array}{cc}
\mathbf{D}\\
\mathbf{B}
\end{array}
\right) =
\left(
\begin{array}{cc}
\epsilon_0\epsilon&i\kappa/c_0\\
-i\kappa/c_0&\mu_0\mu
\end{array}
\right)
\left(
\begin{array}{cc}
\mathbf{E}\\
\mathbf{H}
\end{array}
\right),
\end{equation}
where $\epsilon_0$ and $\mu_0$ are the permittivity and permeability of vacuum, 
$\epsilon$ and $\mu$ are the relative permittivity and permeability of 
the medium respectively. 
$c_0$ is the speed of light in vacuum.
Assuming all fields to be plane waves, the source free Maxwell equations take the form:
\begin{equation}
\label{Maxwell}
\mathbf{k\times H} = -\omega \mathbf{D},\ \ 
\mathbf{k\times E} = \omega \mathbf{B},
\end{equation}
where $\mathbf{k}$ and $\omega$ are the wavevector and the angular frequency of 
the plane wave in the chiral medium. 
Inserting (\ref{constitutive}) into (\ref{Maxwell}), we can get the eigenfunction 
of the electric field $\mathbf{E}$ propagating in the chiral medium:
\begin{equation}
\label{eigenfunction}
\mathbf{k\times(k\times E)} = 
 -k_0^2(\epsilon\mu-\kappa^2)\mathbf{E} - 2i\kappa k_0 (\mathbf{k\times E}),
\end{equation}
where $k_0=\omega/c_0$ is the free space wavevector. 
For simplicity but without loss of generality we can assume $\mathbf{k}=k\hat{z}$
and obtain the eigenvectors and eigenvalues as:
\begin{equation}
\label{eigenvectors}
\mathbf{E}_\pm(z) = (\hat{x}\pm i \hat{y}) E_0e^{ikz},
\end{equation}
\begin{equation}
\label{eigenvalues}
k_\pm = k_0(n\pm\kappa),
\end{equation}
where $n=\sqrt{\epsilon\mu}$. 
There are two eigenmodes in chiral medium. 
One is the right circular polarization (RCP/+) wave and the other is the left 
circular polarization (LCP/-) wave. 
Any wave propagating though the medium can be uniquely decomposed into RCP and LCP and 
there is no conversion between RCP and LCP as they propagate in the chiral medium. 
Then we can define the index of refraction for RCP/LCP waves as:
\begin{equation}
\label{npm}
n_\pm = n\pm\kappa.
\end{equation}
If $\kappa>n$, $n_-$ will be negative. 
This constitutes an alternative approach to realize a negative refractive index 
as proposed earlier by Pendry\cite{pendry}.

A chiral medium has two important properties. One is called optical activity which characterizes the rotation of the polarization plane 
of a linearly polarized light as it passes through a chiral medium. 
Mathematically, it's defined as the polarization azimuth rotation angle of 
elliptically polarized light:
\begin{equation}
\label{theta}
\theta = \frac{1}{2}[\arg(T_+)-\arg(T_-)],
\end{equation}
where $T_+$ and $T_-$ are the transmission coefficients for RCP and LCP waves. 
The other property is circular dichromism which arises from the different absorption
for RCP and LCP. 
It characterizes the difference between the transmissions of two polarizations:
\begin{equation}
\label{eta}
\eta = \frac{1}{2}\tan^{-1}(\frac{|T_+|^2-|T_-|^2}{|T_+|^2+|T_-|^2}).
\end{equation}
Artificial CMMs with large $\theta$ and small $\eta$  
are very important for applications.

% ----------------------------------------------------------------------------
%
\subsection{Parameter retrieval for CMM slabs without substrate}
\label{retrieval for standalone}
%
% Fig. 1
\begin{figure}[htb!]
 \centering{\includegraphics[angle=0, width=8.cm]{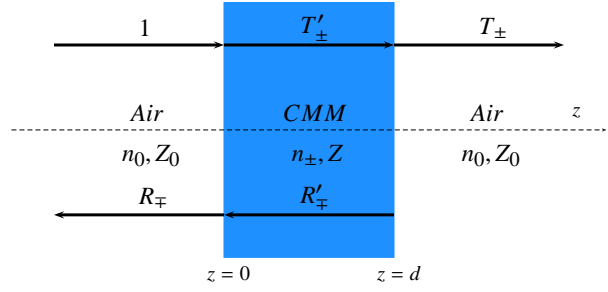}}
 \caption{Schematics of the transmission and reflection coefficient 
  of a standalone CMM slab.}
 \label{standaloneCMM}
\end{figure}
A slab of reciprocal (Pasteur) chiral material looks exactly identical 
for waves propagating in forward or backward direction. However, for a normal incident circular polarized wave
$\mathbf{E}_\pm(z)=(\hat{x}\pm i \hat{y})E_0e^{ikz}$ along the $z$ direction
the reflected wave will be 
$\mathbf{E}^\prime_\pm(z)=(\hat{x}\pm i \hat{y})E_0^\prime\, e^{-ikz}$ 
along the $-z$ direction. 
The $y$ component is always $90^\circ$ advanced(+) / retarded(-) with respect to 
the $x$ component when the wave vector changes to the opposite direction. 
Therefore, the polarization of the reflected wave is reversed based on our current 
definition of handiness in relation to the wave vector, 
i.e.\@ the reflected wave of RCP (LCP) wave is LCP (RCP). 
Now we consider a circularly polarized plane wave normally incidents onto a CMM slab 
with thickness $d$, refractive index $n_\pm$, and impedance $Z=\sqrt{{\mu}/{\epsilon}}$,
as shown in Fig. \ref{standaloneCMM} 
(note that the impedance, $Z$'s, of RCP and LCP  waves are identical \cite{lindell}). 
The amplitude of the incident wave is assumed to be unity; 
the amplitudes of the transmitted and reflected waves are  $T_\pm$ and $R_\mp$ respectively; 
those of the forward and backward propagating waves inside the slab are 
$T_\pm^\prime$ and $R_\mp^\prime$. 
Applying the condition of continuity of tangential electric and magnetic fields at 
$z=0$ and $z=d$, we have the following equations:
\begin{subequations}
\label{boundary}
\begin{align}
&1+R_\mp=T_\pm^\prime+R_\mp^\prime,\\
&1-R_\mp=\frac{T_\pm^\prime-R_\mp^\prime}{Z},\\
&T_\pm^\prime e^{ik_\pm d}+R_\mp^\prime e^{-ik_\mp d}=T_\pm,\\
&\frac{T_\pm^\prime e^{ik_\pm d}-R_\mp^\prime e^{-ik_\mp d}}{Z}=T_\pm,
\end{align}
\end{subequations}
where $k_\pm$ is the wave vector as defined by Eq. (\ref{eigenvalues}). 
Then eliminating $T^\prime$ and $R^\prime$ in Eqs. (\ref{boundary}), we obtain 
the transmission and reflection coefficients:
\begin{subequations}
\label{TR}
\begin{align}
T_\pm &= \frac{4Ze^{ink_0d}e^{\pm i\kappa k_0d}}{(1+Z)^2-(1-Z)^2e^{2ink_0d}}, \\
R_\mp &= \frac{(1-Z^2)(e^{2ink_0d}-1)}{(1+Z)^2-(1-Z)^2e^{2ink_0d}}.
\end{align}
\end{subequations}
Eq. (\ref{TR}b) shows that $R_+ = R_-$. 
This can be understood intuitively as follows: 
We know that the polarizations of  RCP and LCP are reversed after being reflected. 
For the forward incident wave, the optical path of RCP (LCP) is $n_+d$ ($n_-d$), 
where $d$ is the distance that the wave has traveled; 
For the backward reflected wave, the optical path is  $n_-d$ ($n_+d$). 
Therefore, the total optical paths for RCP and LCP are all $n_+d+n_-d$. 
They pass through the same optical path for the reflected waves, so the loss should be 
the same for both of them. 
In addition, the same impedance of the RCP and the LCP gives the same reflections 
for both circular polarizations on the surface. 
Therefore, both the reflections of the single slab should be the same. 

Inverting the above Eqs. (\ref{TR}), 
the impedance $Z$ and refractive index $n_\pm$ are given below:
\begin{subequations}
\label{impedance}
\begin{equation}\label{impedancea}
Z=\pm\sqrt{\frac{(1+R)^2-T_+T_-}{(1-R)^2-T_+T_-}},
\end{equation}
\begin{equation}\label{impedanceb}
n_\pm=\frac{i}{k_0d}\{\ln[\frac{1}{T_\pm}(1-\frac{Z-1}{Z+1}R)]\pm2m\pi\},
\end{equation}
\end{subequations}
where $m$ is an integer determined by the branches. 
The correct sign of the square root in Eq. (\ref{impedancea}) and 
the correct branch of the logarithm in Eq. (\ref{impedanceb}) 
are selected according to the following conditions:
\begin{equation}
\label{restrict}
\mathrm{Re}(Z)\geq0, \mathrm{Im}(n)\geq0,
\end{equation}
which are required by energy conservation and causality\cite{retrievalref_Smith_1}. 
Then the remaining retrieval parameters can be obtained with the following relations: 
$n=(n_++n_-)/2,~\kappa=(n_+-n_-)/2,~\epsilon=n/Z$, and $\mu=nZ$.

Actually, we have another route to get the retrieval results. If we redefine $T=\sqrt{T_+T_-}$ and $R=R_\mp$, according to  Eqs. (\ref{TR}), $T$ and $R$ can be written as:  
\begin{subequations}
\label{TR_linear}
\begin{align}
T&= \frac{4Ze^{ink_0d}}{(1+Z)^2-(1-Z)^2e^{2ink_0d}}, \\
R&= \frac{(1-Z^2)(e^{2ink_0d}-1)}{(1+Z)^2-(1-Z)^2e^{2ink_0d}}.
\end{align}
\end{subequations}
These expressions are exactly the same as those for the ordinary nonchiral materials. Therefore, we can use the traditional retrieval procedure for the nonchiral metamaterials to get the refraction index $n$, the impedance $Z$, the permeability $\epsilon$ and the permittivity $\mu$   \cite{retrievalref_Smith_1}. Then from Eqs. (\ref{TR}a), we can obtain
\begin{equation}
\label{kappafromT}
\kappa=\frac{-i}{2k_0d}\ln(\frac{T_+}{T_-})=\frac{-i}{2k_0d}\ln(\frac{|T_+|e^{i\phi_+}}{|T_-|e^{i\phi_-}}),
\end{equation}
where $\phi_\pm$ are the phases of $T_\pm$. Then we immediately get the refraction indexes $n_\pm=n\pm\kappa$. 

From Eqs. (\ref{kappafromT}), we have
\begin{subequations}
\label{kapparealimag}
\begin{align}
\mathrm{Re}(\kappa)&= \frac{\phi_+-\phi_-+2m\pi}{2k_0d}, \\
\mathrm{Im}(\kappa)&= \frac{\ln|T_-|-\ln|T_+|}{2k_0d},
\end{align}
\end{subequations}
where the integer $m$ is determined by the condition of $-\pi<\phi_+-\phi_-+2m\pi<\pi$ for one unit cell. 
We note that the real and imaginary parts of $\kappa$ relate to the azimuth rotation angle $\theta$ and the circular dichromism $\eta$ respectively. 
%
% ----------------------------------------------------------------------------
%
\subsection{Parameter retrieval for CMM slabs with substrate}
\label{retrieval for substrate}
The retrieval procedure discussed above identifies the scattering amplitudes of 
the CMM slab with those of an equivalent homogeneous slab. 
The retrieved effective parameters of the CMM are the parameters of this equivalent 
homogeneous slab and given as functions of the scattering matrix of the CMM.
In this form, the retrieval procedure is only applicable if the CMM has the same 
symmetry as the equivalent homogeneous slab.
For instance, the reflection amplitudes for a homogeneous chiral (Pasteur) slab 
are identical for incidence from the front or back interfaces. 
For a CMM on top of a substrate (for an example see Fig 2) the scattering amplitudes
do not necessarily have this property. 
In this case, we cannot directly apply the above retrieval procedure to the measured 
or simulated scattering amplitudes of the sample to obtain effective parameters.

In the experiments and simulations (for instance, the commercial software 
CST Microwave Studio \cite{CST}), we can easily obtain the total scattering matrix 
$\mathbf{S}_\mathrm{total}$ of the CMM/substrate slab which relates the incoming waves to the 
outgoing (scattered) waves.  
It is well known\cite{wavepropagation} that any scattering matrix $\mathbf{S}$
corresponds to a transfer matrix $\mathbf{M}$ which relates the amplitudes of the 
in- and outgoing waves in front of the slab to those behind the slab:
\begin{equation}
\label{ST}
\mathbf{S}=
\begin{pmatrix}
r & t^\prime\\ t & r^\prime 
\end{pmatrix}; \ \ \ 
\mathbf{M}=
\begin{pmatrix}
t-rr^\prime/t^\prime & r^\prime/t^\prime\\ -r/t^\prime & 1/t^\prime 
\end{pmatrix},
\end{equation}
where $t$ and $r$ are the transmission and reflection amplitudes;
the quantities without (with) prime ($\prime$) refer to wave propagation in
forward (backward) direction.

Now we consider a structure of substrate/CMM/top layer as an example and assume 
the positive (forward) direction to be from the top layer to substrate layer.
Measurement or simulation afford the total scattering matrix 
$\mathbf{S}_\mathrm{total}$ of the stack in terms of total transmission and 
reflection amplitudes. 
Using (\ref{ST}) we obtain the corresponding total transfer matrix $\mathbf{M}_\mathrm{total}$, 
which factorizes into a product of the transfer matrices of the individual 
layers\cite{wavepropagation}:
\begin{equation}
\label{Matrix}
\mathbf{M}_{\mathrm{total}}=
\mathbf{M}_{\mathrm{substrate}}\mathbf{M}_{\mathrm{CMM}}\mathbf{M}_{\mathrm{top layer}}. 
\end{equation}
The homogeneous substrate and top layers are assumed to be known. 
Their transfer matrices, $\mathbf{M}_{\mathrm{substrate}}$ and 
$\mathbf{M}_{\mathrm{top layer}}$, respectively, are just the well known transfer matix 
of a homogeneous slab and can be calculated analytically 
(as the product of the transfer matix across the interface from vacuum into 
the homogenous slab, the diagonal transfer matrix (phase \& attennuation) for the propagation 
within the homogeneous slab, and the inverse of the interface transfer matrix again to step 
out of the slab back into vacuum):
\begin{equation}
\label{M}
\mathbf{M}=
\left[\begin{array}{cc}\frac{Z+1}{2Z}&\frac{Z-1}{2Z}\\
                       \frac{Z-1}{2Z}&\frac{Z+1}{2Z}\end{array}\right]
\left[\begin{array}{cc}e^{ink_0d}&0\\0&e^{-ink_0d}\end{array}\right]
\left[\begin{array}{cc}\frac{1+Z}{2}&\frac{1-Z}{2}\\
                       \frac{1-Z}{2}&\frac{1+Z}{2}\end{array}\right], 
\end{equation}
where the impedance of $Z$, refractive index of $n$, and the thickness of $d$ 
are all known parameters of the substrate or the top layer. 
Note that the transfer matrix of each layer is that of the slab in free space, 
i.e. the interface of two media can be considered as two media separated by a vacuum slab 
with zero thickness. 

Solving the above equation (\ref{Matrix}), 
we can get the effective transfer matrix of the CMM slab standing alone in free space:
\begin{equation}
\label{CMMMatrix}
\mathbf{M}_{\mathrm{CMM}}=
\mathbf{M}_{\mathrm{substrate}}^{-1}
\mathbf{M}_{\mathrm{total}}
\mathbf{M}_{\mathrm{top layer}}^{-1}. 
\end{equation}
Having extracted the effective transfer matrix of the CMM slab alone, 
we can use the relation (\ref{ST}) again to obtain the S parameters of the free standing CMM slab.
Then the retrival procedure given in Sec. \ref{retrieval for standalone} can be applied to
the isolated S parameters of the CMM slab. 
And following this way, we can obtain the retrival parameters of CMMs with substrate 
and/or the top layer. 

%%%%%%%%%%%%%%%%%%%%%%%%%%%%%%%%%%%%%%%%%%%%%%%%%%%%%%%%%%%%%%%%%%%%%%%%%%%%%%%%
\section{Examples of the CMM slabs with and without substrate} \label{examples}
In this section, we will give two examples to demonstrate that a physically meaningful retrieval 
can be obtained from scattering amplitudes measured for the combined system of CMM and substrate/top layer 
and to show how the substrate influences the retrieval parameters of the CMM slab. 
For this purpose, we choose the twisted-crosses CMMs which had been fabricated in the 
optical region by Decker et al.\cite{chiralexperiments_Decker_2009} and the four-folded rotated $\Omega$-particle CMMs which is closest to the standard chiral model.

\subsection{Retrieval results of the twisted-crosses CMM slabs with and without substrate}\label{twisted-crosses}
%
% Fig. 2
\begin{figure}[htb!]
\hspace{0.6cm}
\centering{\begin{overpic}[width=0.25\textwidth]{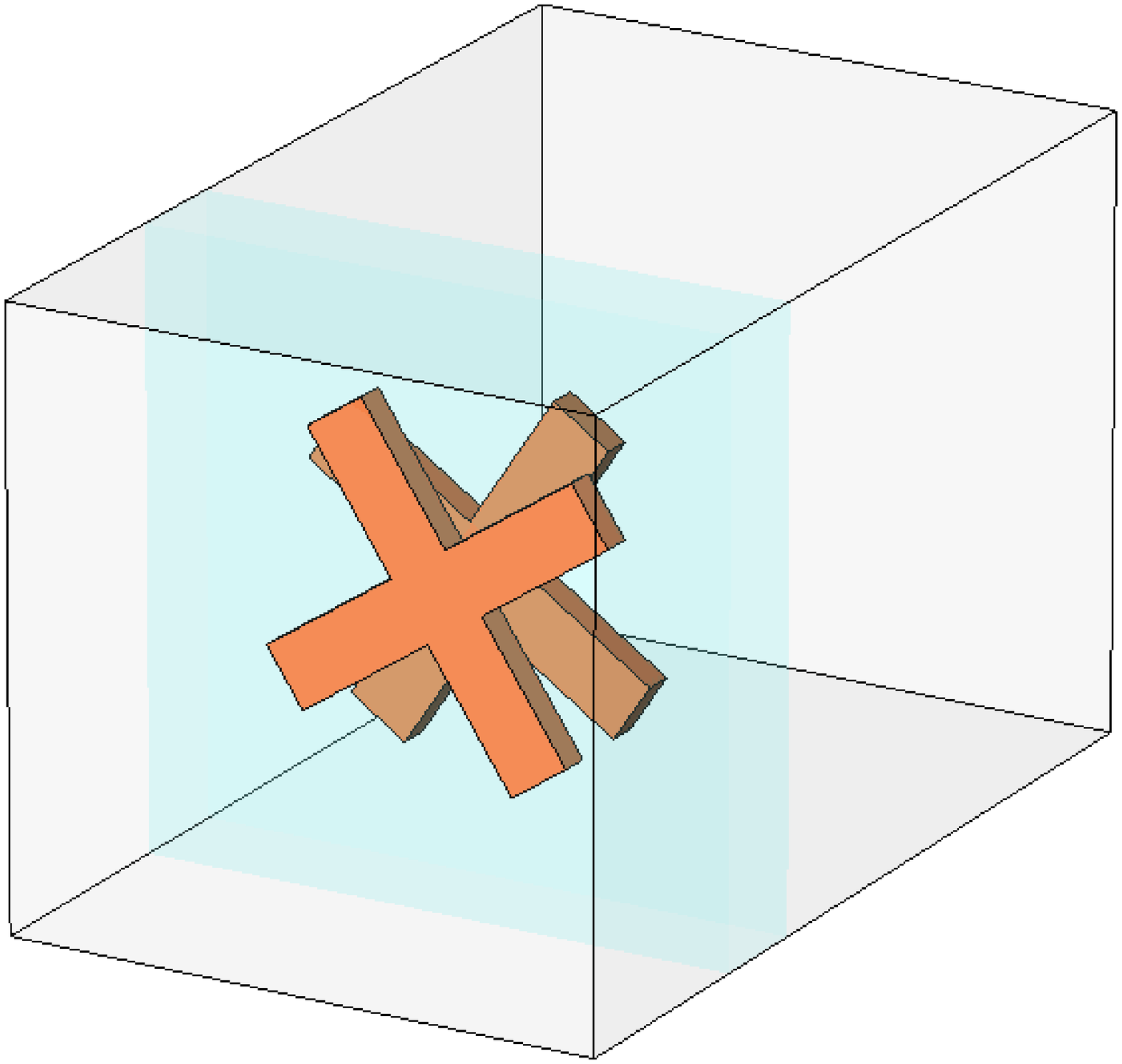}
\put(85,3){\textbf{(a)}} \put(30,100){\scriptsize{\textbf{air/CMM/air}}}
\end{overpic}
%\hspace{0.8cm}
\begin{overpic}[width=0.25\textwidth]{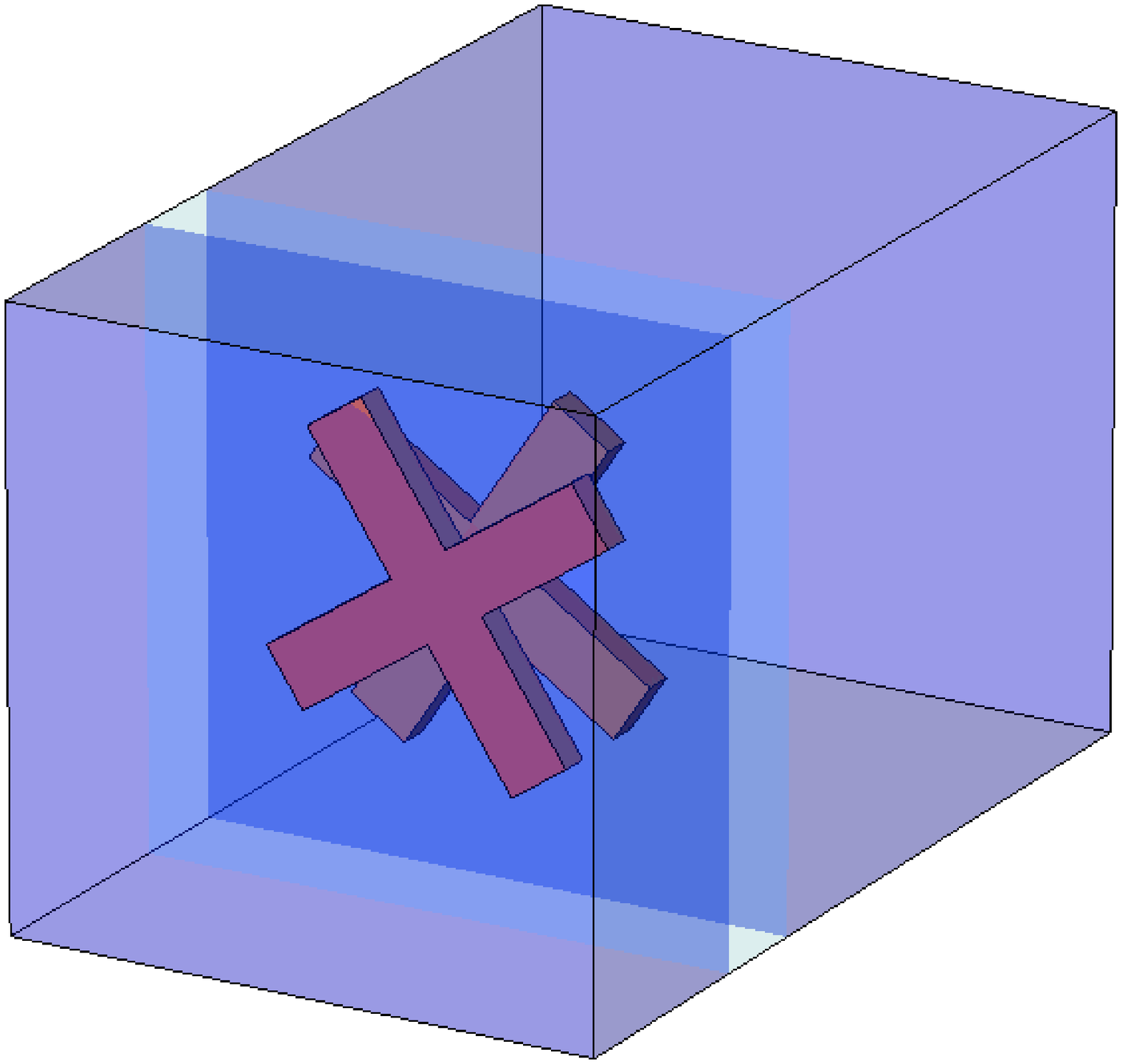}
\put(85,3){\textbf{(b)}}
 \put(0,100){\scriptsize{\textbf{air/top layer/CMM/substrate/air}}}
\end{overpic}
%\hspace{0.8cm}
\begin{overpic}[width=0.25\textwidth]{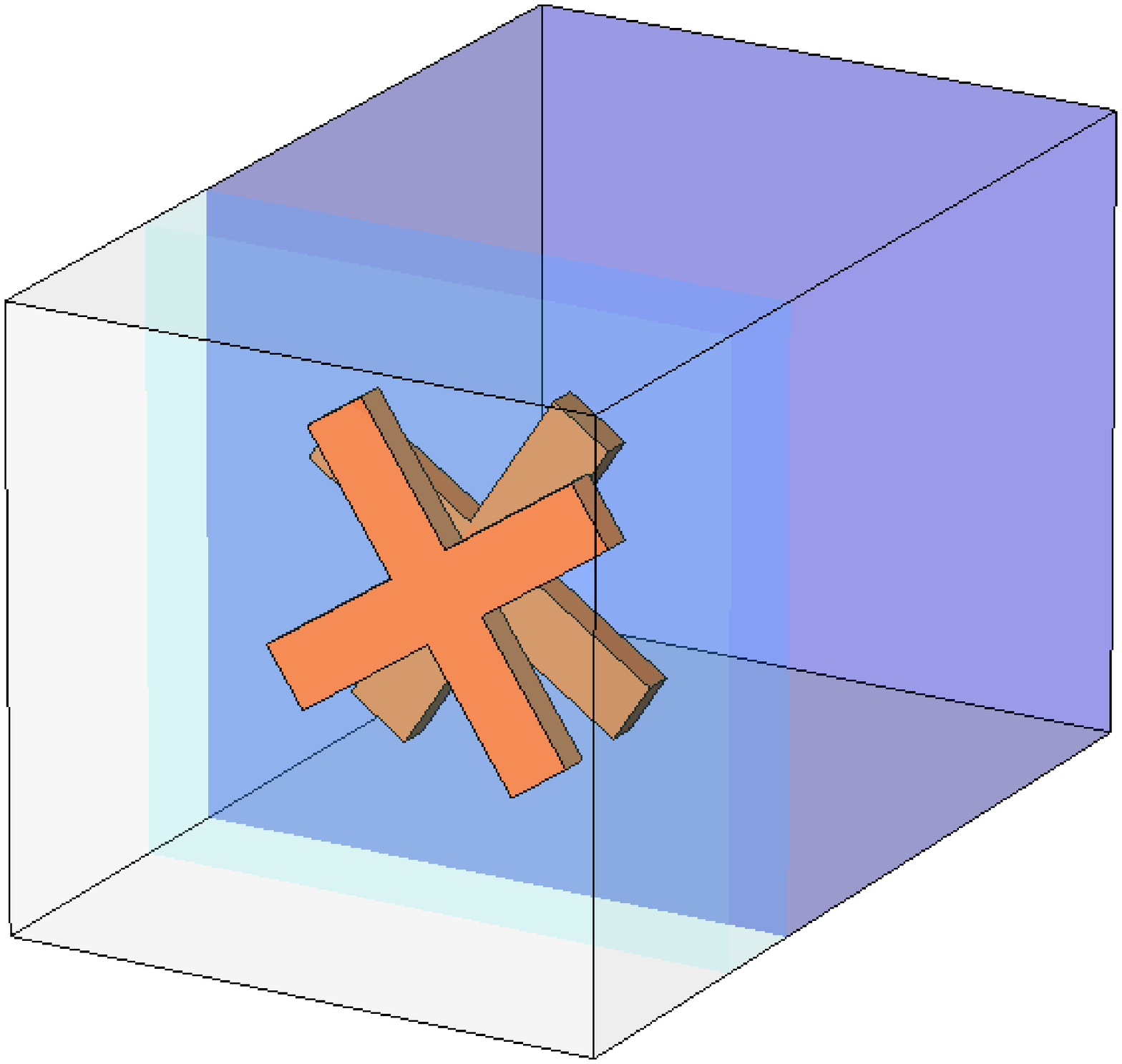}
\put(85,3){\textbf{(c)}}
 \put(15,100){\scriptsize{\textbf{air/CMM/substrate/air}}}
\end{overpic}
\\[2mm]
\begin{overpic}[width=0.87\textwidth]{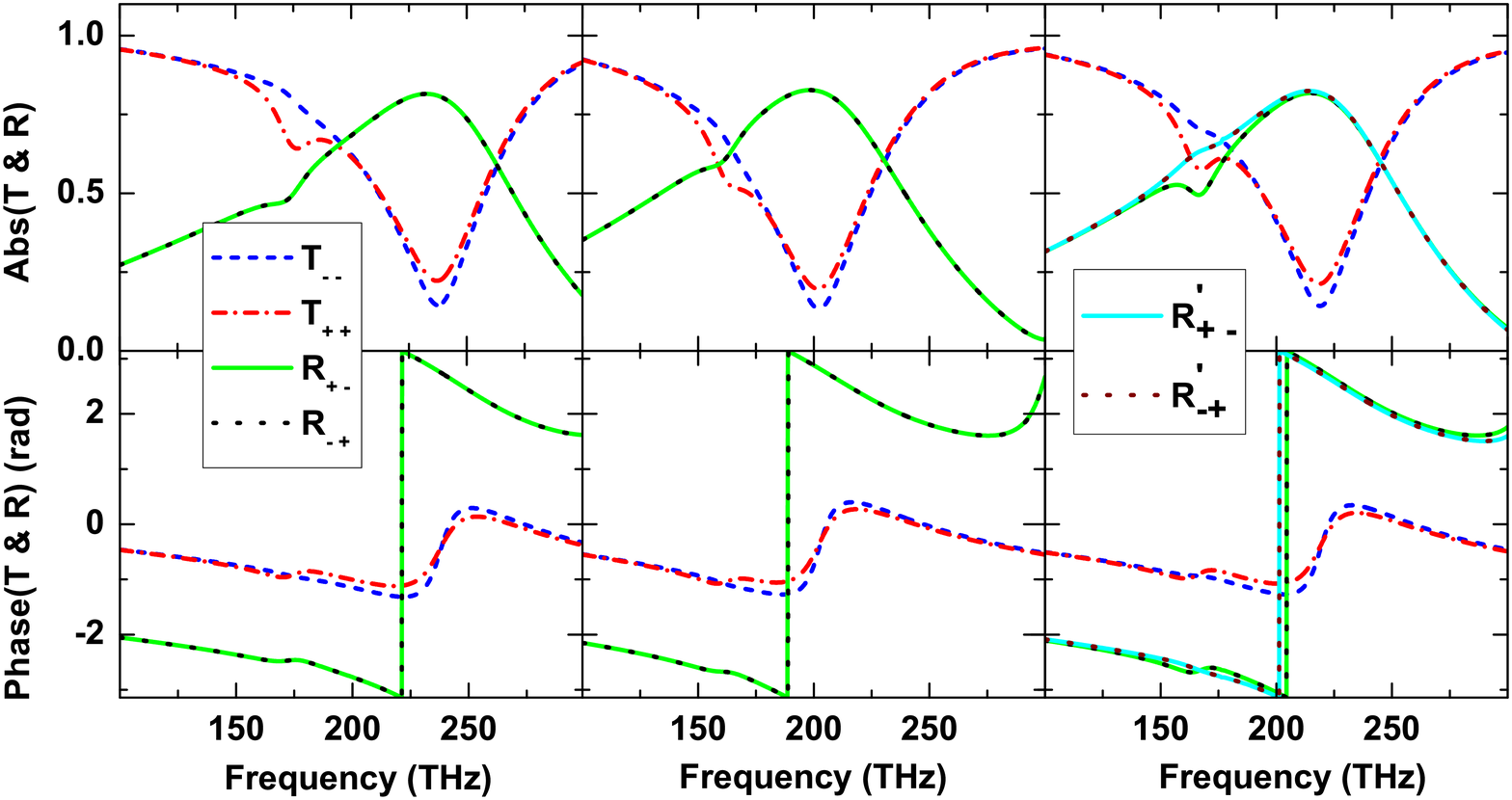}
\put(33.5,30.8){\textbf{(d)}}
\put(63.5,30.8){\textbf{(e)}}
\put(93.5,30.8){\textbf{(f)}}
\put(33.5,8.5){\textbf{(g)}}
\put(63.5,8.5){\textbf{(h)}}
\put(93.5,8.5){\textbf{(i)}}
\end{overpic}}
\caption{(a)-(c) are the Schematics of the twisted-crosses CMM slab. 
 (a) free standing, (b) with substrate and top layer, and (c) with substrate only.
 The amplititudes, (d)-(f),  and the phases, (g)-(i), are also shown correspondingly.
 The prime ($\prime$) denotes that the wave is incident from the opposite direction.
} 
\label{cross_simulation}
\end{figure}

%Fig. 3
\begin{figure}
\centering{\begin{overpic}[width=0.85\textwidth]{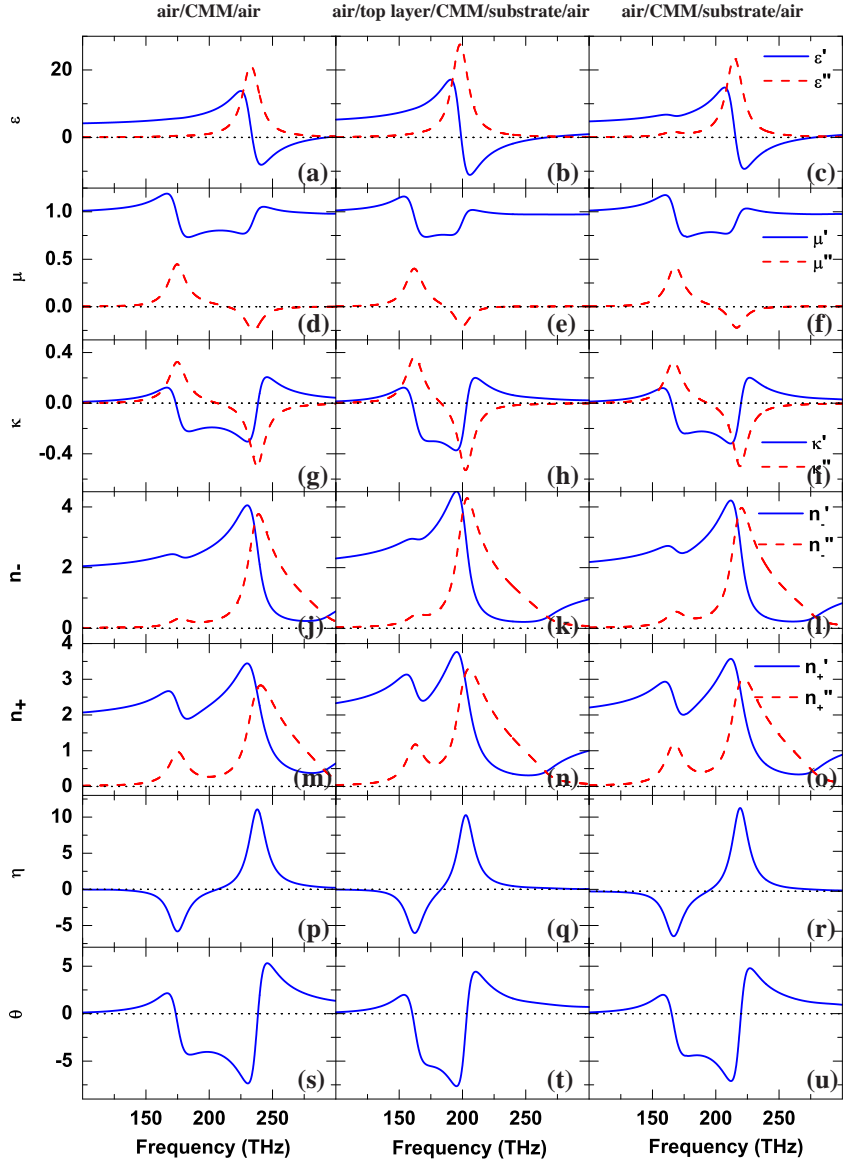}
\put(14,101){\scriptsize{\textbf{air/CMM/air}}}
\put(29.5,101){\scriptsize{\textbf{air/top layer/CMM/substrate/air}}}
\put(55,101){\scriptsize{\textbf{air/CMM/substrate/air}}}

\put(26.4,86.5){\textbf{(a)}}
\put(48.4,86.5){\textbf{(b)}}
\put(71.3,86.5){\textbf{(c)}}

\put(26.4,73.17){\textbf{(d)}}
\put(48.4,73.17){\textbf{(e)}}
\put(71.3,73.17){\textbf{(f)}}

\put(26.4,59.83){\textbf{(g)}}
\put(48.4,59.83){\textbf{(h)}}
\put(71.3,59.83){\textbf{(i)}}

\put(26.4,46.5){\textbf{(j)}}
\put(48.4,46.5){\textbf{(k)}}
\put(71.3,46.5){\textbf{(l)}}

\put(26.0,33.17){\textbf{(m)}}
\put(48.4,33.17){\textbf{(n)}}
\put(71.3,33.17){\textbf{(o)}}

\put(26.4,19.83){\textbf{(p)}}
\put(48.4,19.83){\textbf{(q)}}
\put(71.3,19.83){\textbf{(r)}}

\put(26.4,6.5){\textbf{(s)}}
\put(48.4,6.5){\textbf{(t)}}
\put(71.3,6.5){\textbf{(u)}}

\end{overpic}}
\caption{The retrieval parameters $\epsilon$, $\mu$, $\kappa$, $n_-$, $n_+$, $\eta$, and $\theta$
 for CMM slabs.
 (Left) free-standing, (Middle) with substrate and top layer, and (Right) with substrate only.These results were extracted using (\ref{CMMMatrix}).
} 
\label{cross retrieval}
\end{figure}

The structure of the twisted-crosses CMM is shown in Fig. \ref{cross_simulation}. 
It's composed of two gold crosses rotated against each other by an angle of $22.5^\circ$. 
Each arm of the crosses is a short rectangular wire with the dimension of $25\,\mathrm{nm}\times56\,\mathrm{nm}\times315\,\mathrm{nm}$. 
The separation between the two crosses is $37.5\,\mathrm{nm}$. 
The gold structure is embedded in a lossless dielectric with the refractive index
of $n=1.41$ as shown in Fig. \ref{cross_simulation}(a). 
The structure is fabricated on the glass substrate (n=1.45) with the thickness of $500\,\mathrm{nm}$ as illustrated
in Fig. \ref{cross_simulation}(c); 
Optionally a thin glass layer (n=1.45) with the thickness of $200\,\mathrm{nm}$ can be placed on top of the embedded gold structure, 
see Fig. \ref{cross_simulation}(b).
The gold is modeled by a Drude model permittivity
$\epsilon(\omega)=\epsilon_\infty-\frac{\omega_p^2}{\omega(\omega+i\omega_c)}$ 
with plasma frequency $\omega_p=2\pi\times2159\,\mathrm{THz}$, 
collision frequency $\omega_c=2\pi\times25\,\mathrm{THz}$, 
and $\epsilon_\infty=9.07$. 

The lower two rows in Fig. \ref{cross_simulation} show the circular polarization 
scattering amplitudes (magnitudes and phases) of the CMM slab obtained from simulation.
For the free standing slab (left column) the simulation directly yields the 
scattering amplitudes of the CMM slab;
for the CMM on substrate (right) and CMM on substrat with top layer (middle column) 
the shown CMM scattering amplitudes were extracted using (\ref{CMMMatrix}) derived above. 
$T_{--}$ ($R_{+-}$) and $T_{++}$ ($R_{-+}$) are the transmission (reflection) 
of LCP and RCP, respectively. 
(In order to specify the polarization conversion, for $T_{ab}$ and $R_{ab}$, 
$b$ refers to the incident wave and $a$ to the transmitted/reflected wave.) 
As can be seen in Figs. \ref{cross_simulation}, the reflection amplitudes for RCP and LCP 
are exactly the same, which confirms the previous analysis.

The scattering amplitudes of the homogeneous chiral slab are the same for propagation 
though the slab in forward or backward direction.
The simulated scattering amplitudes for the free standing CMM shown in 
Figs. \ref{cross_simulation}(a,d,g) also have this property. 
We can directly apply the retrieval procedure described in section \ref{retrieval for standalone}
and obtain the effective medium parameters. 
The results are given in Figs. \ref{cross retrieval}.

For the CMM on susbtrate no meaningful retrieval can be obtained by applying the retrieval 
procedure described in section \ref{retrieval for standalone} directly to the total scattering amplitudes;
The presence of the substrate leads to additional scattering not originating from the CMM. 
It also makes the reflections in forward and backward direction unequal - 
an asymmetry which can never be reproduced by the homogeneous chiral slab.
Only after extracting the scattering amplitudes of the CMM slab itself from the simualtion results
using Eq. (\ref{CMMMatrix}) derived in section \ref{retrieval for substrate} 
we can apply the standard retrieval procedure described in section \ref{retrieval for standalone}
to the extracted CMM scattering amplitudes and obtain reasonable effective medium parameters for the CMM slab.

Even after extracting the CMM scattering amplitudes we still observe some weakly asymmetric reflections 
$R_{+-}\neq R_{+-}^\prime$ in forward and backward direction for the CMM on substrate in Figs. \ref{cross_simulation}(f,i).
The extraction derived in section \ref{retrieval for substrate} does only deal with the propagating modes in the system.
However, there is also interaction between the CMM slabs surface regions with their environment mediated by evanescent modes.
This interaction makes the surfaces of the CMM slab on substrate unequal and prevents its approximation by 
an equivalent homogenenous effective medium slab. 

There are two possible solutions to this problem: 
First, we could follow the method introduced in Ref. 22 and {\it approximate} an isotropic reflection $R_{av}$ as the geometric
average,
\begin{equation}
\label{average}
R_{av}=\sqrt{RR^\prime,}
\end{equation}
where, $R$ ($R^\prime$) is the reflection when the incident wave is from
the top/vacuum (bottom/substrate) side of the CMM slab.
This method has been used successfully if the anisotropy was small enough\cite{retrievalref_Smith_2}.
(Note that the inhomogenization is not inherent property of the twisted-crosses CMM slab.
It's only the boundary effect of the thin layer sample.
If a sufficiently transparent sample goes to bulk material with infinity thickness,
this boundary effect becomes negligible.)
Using this method we obtain the retrieval of the single sided CMM on sustrate shown in the right column in Fig. \ref{cross retrieval}.

Second, we could cover the top side of the CMM with a second layer of substrate.
This layer does not need to be very thick as the interaction via the evanescent modes with the surface layers of the CMM decays quickly.
As can be seen in Figs. \ref{cross_simulation}(e,h) this approach lead to perfectly symmetric reflection amplitudes and allows to apply
the standard retrieval procedure described in section \ref{retrieval for standalone}
to the extracted CMM scattering amplitudes to obtain the effective parametes of the CMM slab.
The results for this case are show in the center column in Fig. \ref{cross retrieval}.

Comparing the results with and without substrate, we can see that the substrate has two main influences on the CMMs.  One is lowering 
the frequency of the resonance because the substrate increases the capacity $C$ of the resonator and $\omega_R\sim1/\sqrt{LC}$. The other is the substrate may induce the homogeneous CMM to be inhomogeneous.  For the structure with substrate only, the two crosses are in the different environments. For the cross near the substrate, the evanescent field leak to the substrate while for the cross near the air, the field leaks to the air, which makes the two substantially identical crosses unequal physically. Even the substrate contribution has been deducted off mathematically 
by using Eq. (\ref{CMMMatrix}), 
the CMM slab still shows the inhomogeneous property (e.q. $R_{+-}\neq R_{+-}^\prime$). 

% ----------------------------------------------------------------------------
\subsection{Retrieval results of the four-folded rotated $\Omega$-particle CMM slabs with and without substrate}\label{Omega particle}
%Fig. 4
\begin{figure}[htb!]
 \centering{\includegraphics[angle=0, width=0.5\textwidth]{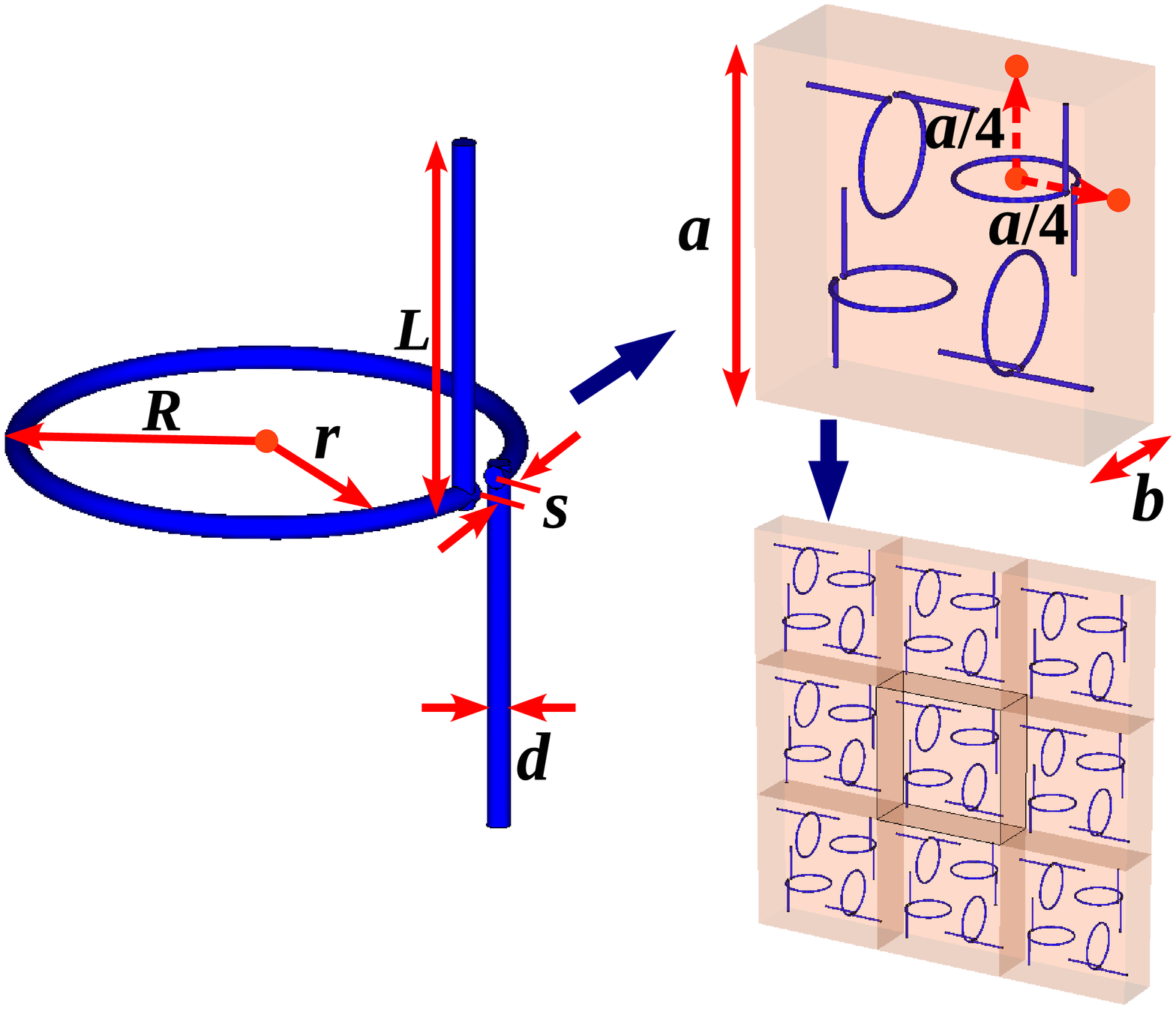}}
\caption{Schmetics of the four-folded rotated $\Omega$-particle CMM. The inner and
outer radius of the loop are $r=5\,\mathrm{\mu m}$ and $R=5.5\,\mathrm{\mu m}$. The length of each straight wire is $L=8.08\,\mathrm{\mu m}$. $d=0.5\,\mathrm{\mu m}$. $a=32\,\mathrm{\mu m}$. $b=12\,\mathrm{\mu m}$. Each $\Omega$-particle locates at the center of each quarter of the unit cell. The silver $\Omega$-particle is embeded in the polyimide with $\epsilon=2.5$ and loss tangent $\delta=0.03$. The silver is charactored by the Drude model with the surface plasmon $\omega_p=13.66\times 10^{15} \,\mathrm{rad/s}$ and the collision frequency $\omega_c=2.73\times10^{13}\,\mathrm{rad/s}$.}
 \label{omega}
\end{figure}
Not all the chiral metamaterials are so sensitive to the influence of the substrate.
In the following, we will give such an example---four-folded
rotated $\Omega$-particle chiral metamaterials. The structure and the dimensional parameters are shown in Fig. \ref{omega}. The incident wave is perpendicular to the CMM slab. 
% Fig. 5
\begin{figure}[htb!]
 \centering{\begin{overpic}[width=0.6\textwidth]{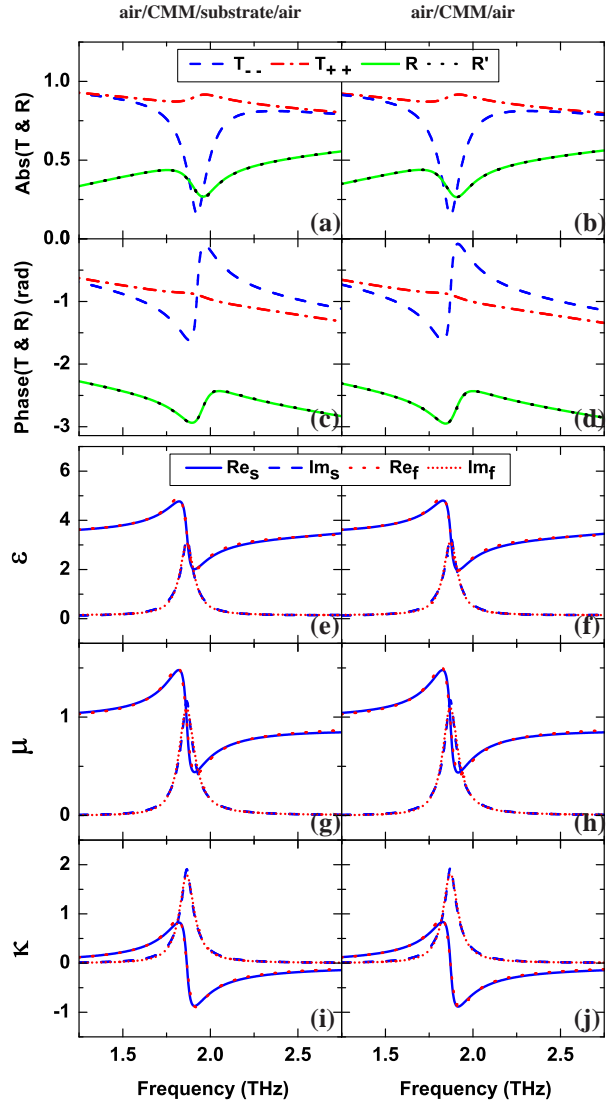}
\put(10,97){\scriptsize{\textbf{air/CMM/substrate/air}}}
\put(36,97){\scriptsize{\textbf{air/CMM/air}}}

\put(27,78){\textbf{(a)}}
\put(50.5,78){\textbf{(b)}}

\put(27,60.5){\textbf{(c)}}
\put(50.5,60.5){\textbf{(d)}}

\put(27,42){\textbf{(e)}}
\put(50.5,42){\textbf{(f)}}

\put(27,24.5){\textbf{(g)}}
\put(50.5,24.5){\textbf{(h)}}

\put(27,7){\textbf{(i)}}
\put(50.5,7){\textbf{(j)}}

\end{overpic}}

 \caption{(a)-(d) are the simulation results of the amplitude and the phase
  of the reflection and transmission for the four-folded
rotated $\Omega$-particle CMM slabs. They are extracted using (\ref{CMMMatrix}). $R^\prime$ is the reflection for the incident wave from the opposite direction.
  (e)-(j) show the optical parameters $\epsilon$, $\mu$, and $\kappa$. $\mathrm{Re}$/$\mathrm{Im}$ denotes the real/imaginary part of the value. The subscripts $\mathrm{s}$ and $\mathrm{f}$ denotes the retrieval results from the numerical simulation and the fitting results using Eqs. (\ref{emk}) respectively. The left is with substrate (n=1.27) only and the right is for the free standing sample.}
 \label{omegaresults}
\end{figure}

The amplitude and phase of the reflection and transmission are
shown in Figs. \ref{omegaresults}(a-d). The left is with $50\,\mathrm{\mu m}$ substrate (n=1.27) only and the right is for the free standing sample.
There is a strong resonance at around $1.87\,\mathrm{THz}$ where both the amplitude 
and phase of the transmissions of RCP and LCP are different, 
which indicates very strong optical activity and circular dichromism. 
The interesting is that for the structure with substrate only, the extracted reflections from different directions almost overlap with each other, e.g. $R=R^\prime$, even they have a large difference before the extraction using Eq. (\ref{CMMMatrix}).  Comparing the results between with (Fig. \ref{omegaresults}(a,c)) and without (Fig. \ref{omegaresults}(b,d)) substrate, we see that the substrate almost has no influence on the $\Omega$-particle CMMs. That is because the resonance elements, the gap of the circle magnetic resonator and the short wires for the electric resonance, are located in the middle of the unit cell. They are far away from the unit cell surface, in other words, they are far away from the substrate. The evanescent field leaked into the substrate is very little. Therefore, this kind CMM is not sensitive to the existence of the substrate.

Figs. \ref{omegaresults}(e-j) are the retrieval results of the optical parameters  $\epsilon$, $\mu$, and $\kappa$. All these curves behave well. We fit these curves using the analytical formulas given by Eqs. (\ref{emk}) derived from the $\Omega$-particle resonator model (see Appendix \ref{appendix}). For with substrate only, the fitted parameters are as following: 
$\epsilon_b=3.1738,~
 \mu_b=0.9799,~
 \omega_0=1.8651\,\mathrm{THz},~
 \gamma=0.05519\,\omega_0,~
 \Omega_\epsilon=0.1537,~
 \Omega_\mu=0.0627,$ and 
$\Omega_\kappa=0.0986$. 
For without substrate, they are very close to those with substrate. 
$\epsilon_b=3.1736,~
 \mu_b=0.9798,~
 \omega_0=1.8713\,\mathrm{THz},~
 \gamma=0.05463\,\omega_0,~
 \Omega_\epsilon=0.1560,~
 \Omega_\mu=0.0625,$ and 
$\Omega_\kappa=0.0993$. From these parameters, we can again see that the substrate has very little influence on the four-folded
rotated $\Omega$-particle CMMs. The substrate only lowers the frequency of the resonance by 0.3\%.

In Figs. \ref{omegaresults}(e-j), we can also see that the retrieval results from numerical simulations agree very well with the 
analytical results given by Eqs. (\ref{emk}). It verifies that the $\omega$ dependences of the optical parameters are valid in our previous work about realizing the repulsive Casimir force \cite{rkzhao} using chiral metamaterials. Note that from Eq. (\ref{pm1}), the strength of electric polarization and 
magnetic polarization can be modified independently by the length of the straight wires 
and the area of the circular loop respectively. $\Omega_\epsilon>\Omega_\mu$ here is just one case. But no matter how to change the dimensional parameters, 
the relation Eq. (\ref{KME}) roughly always holds in such designs composited by the passive material only.

%%%%%%%%%%%%%%%%%%%%%%%%%%%%%%%%%%%%%%%%%%%%%%%%%%%%%%%%%%%%%%%%%%%%%%%%%%%%%%%
%
\section{Conclusion}
\label{conclusion}
In summary, we have analyzed the wave propagation in chiral metamaterials and followed 
the similar procedure that was done in ordinary metamaterials to do the parameter retrieval 
for CMM slabs. 
Based on the transfer matrix technique, we extended the parameter retrieval technique 
to be able to treat samples with substrates and extra top layers. We studied the influence of the substrate and top layers on the thin chiral metamaterial slabs. We found that the substrate could lower the frequency of the resonance and induce the homogeneous chiral metamaterials to be inhomogeneous. The sensitivity of the influence relates to the strength of the leaked evanescent field in the substrate. We also fitted the retrieval results using analytical expressions derived from $\Omega$-particle chiral model based on the effective LC circuit mode. We found they agree with each other very well.

%%%%%%%%%%%%%%%%%%%%%%%%%%%%%%%%%%%%%%%%%%%%%%%%%%%%%%%%%%%%%%%%%%%%%%%%%%%%%%%%%
%\begin{acknowledgements}
\section*{Acknowledgements}
Work at Ames Laboratory was supported by the Department of Energy
(Basic Energy Sciences) under contract No.~DE-AC02-07CH11358. This
work was partially supported by the European Community FET project
PHOME (contract No.~213390). The author Rongkuo Zhao
specially acknowledges the China Scholarship Council (CSC) for financial support.
%\end{acknowledgements}
%%%%%%%%%%%%%%%%%%%%%%%%%%%%%%%%%%%%%%%%%%%%%%%%%%%%%%%%%%%%%%%%%%%%%%%%%%%%%%%%%%
%
\appendix
\section{Derive the forms of the constitutive parameters from the $\Omega$-particle resonator model}\label{appendix}
\begin{figure}[htb!]
 \centering{\includegraphics[angle=0, width=0.5\textwidth]{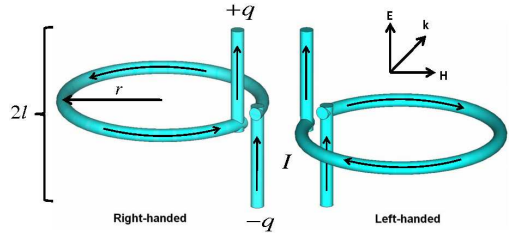}}
 \caption{Schematics of the single right-handed (left) and left-handed (right) $\Omega$-particle resonators.}
 \label{helix}
\end{figure}
Fig. \ref{helix} gives the schematics of the $\Omega$-particle
resonator. The structure consists of an open circular loop and two
short wires. The area of the loop is $S=\pi r^2$ and the length of
each wire is $l$. The wires are perpendicular to the loop and
connected to the ends of the open loop. If the $\Omega$-particle is
in a homogeneous external field, the driving electric potential can
be written as:
\begin{equation}
U=2lE_0\pm\mu_0S\dot{H}_0,
\end{equation}
where $\pm$ signs correspond to right-handed and left-handed helix
resonators, The hat-dot denotes the first time derivative. Applying
the effective RLC circuit model, we have
\begin{equation}
L \dot{I}+\frac{q}{C}+RI=U,\ \ \ \ I=\dot{q}.
\end{equation}
Let $\alpha=\frac{l}{L}$, $\beta=\mu_0\frac{A}{L}$, $\gamma=\frac{R}{L}$, and
$\omega^2_0=\frac{1}{LC}$. Then assuming the external fields are
plane waves, $\sim e^{-i\omega t}$, and considering stationary
solutions, we get the following equation:
\begin{equation}
(-\omega^2-i\omega\gamma+\omega_0^2)q=\alpha E_0\pm i\omega\beta H_0,
\end{equation}
then, we have 
\begin{equation}
q=\frac{\alpha}
       {-\omega^2-i\omega\gamma+\omega_0^2}
  E_0\pm
  \frac{i\omega\beta}
       {-\omega^2-i\omega\gamma+\omega_0^2}
  H_0.
\end{equation}
Therefore, the electric dipole, $\mathbf{p}=q\mathbf{l}$, and the
magnetic dipole, $\mathbf{m}=\pm
I\mathbf{A}=\pm\dot{q}\mathbf{A}$, can be written as:
\begin{subequations}
\label{pm1}
\begin{eqnarray}
\mathbf{p} &= \frac{\alpha\mathbf{l}}
                   {\omega_0^2-\omega^2-i\omega\gamma}
              E_0 +
              \frac{\pm i\omega\beta\mathbf{l}}
                   {\omega_0^2-\omega^2-i\omega\gamma}
              H_0, \\
\mathbf{m} &= \frac{\mp i\omega\alpha\mathbf{A}}
                   {\omega_0^2-\omega^2-i\omega\gamma}
              E_0 +
              \frac{\omega^2\beta\mathbf{A}}
                   {\omega_0^2-\omega^2-i\omega\gamma}
              H_0,
\end{eqnarray}
\end{subequations}
where, $\alpha\mathbf{A}=\frac{1}{\mu_0}\beta\mathbf{l}$,
$\mathbf{l}=l\hat{\mathbf{l}}_0$, and $\mathbf{A}=A\hat{\mathbf{l}}_0$; 
$\hat{\mathbf{l}}_0$ is the unit vector along the direction of the wires, 
i.e., the up direction. 
The electric polarization and the magnetic polarization are defined as  
$\mathbf{P}=\frac{\sum \mathbf{p}}{V_0}$ and 
$\mathbf{M}=\frac{\sum \mathbf{m}}{V_0}$ respectively, 
where the summation is done in the unit cell and $V_0$ denotes the volume of the unit cell. 
Assuming there are $N$ resonators in one unit cell, 
the electric polarization and the magnetic polarization can be written as:
\begin{subequations}
\label{PM}
\begin{eqnarray}
\mathbf{P} &= \frac{N}{V_0}
              \frac{\alpha l}{\omega_0^2-\omega^2-i\omega\gamma} \mathbf{E} + 
              \frac{N}{V_0}
              \frac{\pm i\omega\beta l}{\omega_0^2-\omega^2-i\omega\gamma} \mathbf{H},
              \\
\mathbf{M} &= \frac{N}{V_0}
              \frac{\mp i\omega\alpha A}{\omega_0^2-\omega^2-i\omega\gamma} \mathbf{E} +
              \frac{N}{V_0}
              \frac{\omega^2\beta A}{\omega_0^2-\omega^2-i\omega\gamma} \mathbf{H}.
\end{eqnarray}
\end{subequations}
Here, the directions of $\mathbf{l}$ and $\mathbf{A}$ are merged into $E_0$ and $H_0$ 
to form the vectors of $\mathbf{E}$ and $\mathbf{H}$. 

Inserting Eqs. (\ref{PM}) to the relation 
$\mathbf{D}=\epsilon_0\mathbf{E}+\mathbf{P}$ and 
$\mathbf{B}=\mu_0\mathbf{H}+\mu_0\mathbf{M}$, we have
\begin{subequations}
\label{DB}
\begin{eqnarray}
\mathbf{D}\!\!\! &=&
 \!\!\!\epsilon_0\mathbf{E} \!+\!
 \frac{\alpha l N/V_0}{\omega_0^2-\omega^2-i\omega\gamma} \mathbf{E} \!+\!
 \frac{\pm i\omega\beta l N/V_0}{\omega_0^2-\omega^2-i\omega\gamma} \mathbf{H},
 \\
\mathbf{B}\!\!\! &=& 
 \!\!\!\mu_0\mathbf{H} \!+\! 
 \frac{\mp i\mu_0 \omega\alpha A N/V_0}{\omega_0^2-\omega^2-i\omega\gamma} \mathbf{E} \!+\!
 \frac{\omega^2\mu_0\beta A N/V_0}{\omega_0^2-\omega^2-i\omega\gamma} \mathbf{H}.
\end{eqnarray}
\end{subequations}
Comparing Eqs. (\ref{DB}) with Eq. (\ref{constitutive}), 
we can obtain the relative permittivity $\epsilon$, permeability $\mu$, and 
the chirality $\kappa$:
\begin{subequations}
\label{emk_vacuum}
\begin{equation}
\epsilon = 1+\frac{\alpha l N/V_0 \epsilon_0}{\omega_0^2-\omega^2-i\omega\gamma},
\end{equation}
\begin{equation}
\mu = 1+\frac{\omega^2\beta A N/V_0}{\omega_0^2-\omega^2-i\omega\gamma},
\end{equation}
\begin{equation}
\kappa = \frac{\pm \omega\beta l c_0 N/V_0}{\omega_0^2-\omega^2-i\omega\gamma}
       = \frac{\pm \omega \mu_0 c_0 \alpha A N/V_0}{\omega_0^2-\omega^2-i\omega\gamma}.
\end{equation}
\end{subequations}
Eqs. (\ref{emk_vacuum}) are valid for the case that the metal chiral structure 
stands in free space. 
If it's merged into a background material with $\epsilon_b$ and $\mu_b$, 
Eqs. (\ref{emk_vacuum}) should be modified as follows:
\begin{subequations}
\label{emk}
\begin{equation}
\label{emk_e}
\epsilon = \epsilon_b+\frac{\Omega_\epsilon\omega_0^2}{\omega_0^2-\omega^2-i\omega\gamma},
\end{equation}
\begin{equation}
\label{emk_m}
\mu = \mu_b+\frac{\Omega_\mu\omega^2}{\omega_0^2-\omega^2-i\omega\gamma},
\end{equation}
\begin{equation}
\label{emk_k}
\kappa = \frac{\Omega_\kappa\omega_0\omega}{\omega_0^2-\omega^2-i\omega\gamma},
\end{equation}
\end{subequations}
where $\epsilon_b$ is usually larger than one; $\mu_b$ is usually very close to one. 
$\Omega_\epsilon$, $\Omega_\mu$, and $\Omega_\kappa$ are the coefficients of 
the resonance terms in $\epsilon$, $\mu$, and $\kappa$, i.e. 
$\Omega_\epsilon=\frac{\alpha l N}{ V_0 \epsilon_0 \omega_0^2}$, 
$\Omega_\mu=\frac{\beta A N}{V_0}$, and 
$\Omega_\kappa=\frac{\beta l c_0 N}{V_0 \omega_0} =
 \frac{\mu_0 c_0 \alpha A N}{V_0 \omega_0}$, 
where $c_0$ is the speed of light in vacuum. 
They describe the strength of the resonance. The expressions of the frequency dependence of $\epsilon$ and $\mu$ given by 
Eqs. (\ref{emk_e}) and (\ref{emk_m}) are the same as those in traditional 
metamaterials \cite{Pendry_mu}. 
The linear $\omega$ dependence of $\kappa$ is the same as Condon model 
for the homogeneous chiral molecular media \cite{condon}. 
This is a general feature of natural optically active materials \cite{Landau}. 

Because $\alpha=\frac{l}{L}$ and $\beta=\mu_0\frac{A}{L}$, we then
can obtain the following relation:
\begin{equation}\label{KME}
\Omega_\kappa^2=\Omega_\epsilon\Omega_\mu,
\end{equation}
which seems to limit the possibility of obtaining large
$\Omega_\kappa$ in these designs composited by the passive materials only.


\begin{thebibliography}{99}

\bibitem{pendry}
J. B. Pendry, ``A chiral route to negative refraction,'' 
Science \textbf{306}, 1353-1355 (2004).
\
\bibitem{Tretyakov}
S. Tretyakov, I. Nefedov, A. Sihvola, S. Maslovski, and C. Simovski, ``Waves and energy in chiral nihility,''
J. Electromagn. Waves Appl. \textbf{17}, 695-706 (2003).

\bibitem{Monzon}
Cesar Monzon and D. W. Forester, ``Negative refraction and focusing of circularly polarized waves in optically active media,''
Phys. Rev. Lett. \textbf{95}, 123904 (2005).

\bibitem{TretyakovPNFA}
S. Tretyakov, A. Sihvola, and L. Jylha, “Backward-wave regime and negative refraction in chiral composites,”
Photonics Nanostruct. Fundam. Appl. \textbf{3}, 107 (2005).

\bibitem{YannopapasJPCM}
V. Yannopapas, "Negative index of refraction in artificial chiral materials," J. Phys.: Condens. Matter 18, 6883-6890 (2006).

\bibitem{AgranovichPRB}
V. M. Agranovich, Y. N. Gartstein, and A. A. Zakhidov, "Negative refraction in gyrotropic media," Phys. Rev. B 73, 045114 (2006).

\bibitem{negativereflection}
C. Zhang and T. J. Cui, ``Negative reflections of electromagnetic waves in a strong chiral medium,''
Appl. Phys. Lett. \textbf{91}, 194101 (2007).

\bibitem{chiral_Martin}
J. K. Gansel, M. Thiel, M. S. Rill, M. Decker, K. Bade, V. Saile, 
G. V. Freymann, S. Linden, and M. Wegener, ``Gold helix photonic metamaterial as broadband circular polarizer,''
Science \textbf{325}, 1513-1515 (2009).

\bibitem{chiral_Plum}
E. Plum, J. Zhou, J. Dong, V. A. Fedotov, Th. Koschny, 
C. M. Soukoulis, and N. I. Zheludev, ``Metamaterial with negative index due to chirality,''
Phys. Rev. B \textbf{79}, 035407 (2009).

\bibitem{chiral_Zhou}
J. Zhou, J. Dong, B. Wang, Th. Koschny, M. Kafesaki, and C. M. Soukoulis, ``Negative refractive index due to chirality,''
Phys. Rev. B \textbf{79}, 121104(R) (2009).

\bibitem{chiral_Dong}
J. Dong, J. Zhou, Th. Koschny, and C. M. Soukoulis, ``Bi-layer cross chiral structure with strong optical activity and negative refractive index,''
Opt. Express \textbf{17}, 14172-14179 (2009).

\bibitem{chiral_Plum_APL2008}
E. Plum, V. A. Fedotov, and N. I. Zheludev, ``Optical activity in extrinsically chiral metamaterial,''
Appl. Phys. Lett. \textbf{93}, 191911 (2008).

\bibitem{chiral_Plum_PRL2009}
E. Plum, X.-X. Liu, V. A. Fedotov, Y. Chen, D. P. Tsai, and N. I. Zheludev, ``Metamaterials: optical activity without chirality,''
Phys. Rev. Lett. \textbf{102}, 113902 (2009).

\bibitem{chiral_Jelinek}
L. Jelinek, R. Marqu\ifmmode \check{e}\else \v{e}\fi{}s, F. Mesa, and J. D. Baena, ``Periodic arrangements of chiral scatterers providing negative refractive index bi-isotropic media,''
Phys. Rev. B \textbf{77}, 205110 (2008).

\bibitem{chiral_Bingnan}
B. Wang, J. Zhou, Th. Koschny, and C. M. Soukoulis, ``Nonplanar chiral metamaterials with negative index,''
Appl. Phys. Lett. \textbf{94}, 151112 (2009).

\bibitem{chiral_Yannopapas}
V. Yannopapas, ``Circular dichroism in planar nonchiral plasmonic metamaterials,''
Opt. Lett. \textbf{34}, 632-634 (2009).

\bibitem{chiralexperiments_Zhang}
S. Zhang, Y. S. Park, J. Li, X. Lu, W. Zhang, and X. Zhang, ``Negative refractive index in chiral metamaterials,''
Phys. Rev. Lett. \textbf{102}, 023901 (2009).

\bibitem{chiralexperiments_Plum}
E. Plum, V. A. Fedotov, A. S. Schwanecke, Y. Chen, and N. I. Zheludev, ``
Giant optical gyrotropy due to electromagnetic coupling
,'' Appl. Phys. Lett. \textbf{90}, 223113 (2007).

\bibitem{chiralexperiments_Gonokami}
M. Kuwata-Gonokami, N. Saito, Y. Ino, M. Kauranen, K. Jefimovs, 
T. Vallius, J. Turunen, and Y. Svirko, ``Giant optical activity in quasi-two-dimensional planar nanostructures,''
Phys. Rev. Lett. \textbf{95}, 227401 (2005). 

\bibitem{chiralexperiments_Decker_2010}
M. Decker, R. Zhao, C.M. Soukoulis, S. Linden, and M. Wegener, ``Twisted split-ring-resonator photonic metamaterial with huge optical activity,'' Opt. Lett., \textbf{35}, 1593-1595 (2010).

\bibitem{chiralexperiments_Decker_2009}
M. Decker, M. Ruther, C. E. Kriegler, J. Zhou, C. M. Soukoulis, 
S. Linden, and M. Wegener, ``Strong optical activity from twisted-cross photonic metamaterials,''
Opt. Lett. \textbf{34}, 2501-1503 (2009).

\bibitem{chiralexperiments_Decker_2007}
M. Decker, M. W. Klein, M. Wegener, and S. Linden, ``Circular dichroism of planar chiral magnetic metamaterials,''
Opt. Lett. \textbf{32}, 856-858 (2007).

\bibitem{retrievalref_Smith_1}
D. R. Smith, S. Schultz, P. Marko\ifmmode \check{s}\else \v{s}\fi{}, and C. M. Soukoulis, ``Determination of effective permittivity and permeability of metamaterials from reflection and transmission coefficients,''
Phys. Rev. B \textbf{65}, 195104 (2002).

\bibitem{retrievalref_Chen}
X. Chen, T. M. Grzegorczyk, B. I. Wu, J. Pacheco, and J. A. Kong, ``Robust method to retrieve the constitutive effective parameters of metamaterials,''
Phys. Rev. E \textbf{70}, 016608 (2004).

\bibitem{retrievalref_Koschny_1}
Th. Koschny,M. Kafesaki, E. N. Economou, and C. M. Soukoulis, ``Effective medium theory of left-handed materials,''
Phys. Rev. Lett. \textbf{93}, 107402 (2004).

\bibitem{retrievalref_Koschny_2}
Th. Koschny, P. Marko\ifmmode \check{s}\else \v{s}\fi{}, 
E. N. Economou, D. R. Smith, D. C. Vier, and C. M. Soukoulis, ``Impact of inherent periodic structure on effective medium description of left-handed and related metamaterials,''
Phys. Rev. B \textbf{71}, 245105 (2005).

\bibitem{retrievalref_Smith_2}
D. R. Smith, D. C. Vier, Th. Koschny and C. M. Soukoulis, ``Electromagnetic parameter retrieval from inhomogeneous metamaterials,''
Phys. Rev. E \textbf{71}, 036617 (2005).

\bibitem{retrievalref_Li}
Z. Li, K. Aydin, and E. Ozbay, ``Determination of the effective constitutive parameters of bianisotropic metamaterials from reflection and transmission coefficients,''
Phys. Rev. E \textbf{79}, 026610 (2009).

\bibitem{BingnanJOA}
B. Wang, J. Zhou, Th. Koschny, M. Kafesaki, and C. M. Soukoulis, ``Chiral metamaterials: simulations and experiments,'' J. Opt. A: Pure Appl. Opt. \textbf{11}, 114003 (2009).

\bibitem{retrievalwithsubstrate}
D. H. Kwon, D. H. Werner, A. V. Kildishev, and V. M. Shalaev, ``Material parameter retrieval procedure for general bi-isotropic metamaterials and its application to optical chiral negative-index metamaterial design,''
Opt. Express \textbf{16}, 11822-11829 (2008).


\bibitem{lindell}
I. V. Lindell et al., 
\textit{Electromagnetic Waves in Chiral and Bi-Isotropic Media} 
(Artech House, Boston $\cdot$ London, 1994).

\bibitem{Serdyukov}
A. Serdyukov et al., 
\textit{Electromagnetics of Bi-anisotropic Materials: Theory and Applications} 
(Gordon and Breach Science Publishers, Amsterdam, 2001).

\bibitem{constitutive}
Some people 
(see, for instance, A. Lakhtakia et al., ``Reflection of plane waves at planar achiral-chiral interfaces: independence of the reflected polarization state from the incident polarization state,'' J. Opt. Soc. Am. A \textbf{7}, 1654 (1990).) 
use the Drude-Born-Fedorov relations: 
$\mathbf{D} = \epsilon(\mathbf{B}+\beta\nabla\times\mathbf{E}),~
 \mathbf{B} = \mu(\mathbf{H}+\beta\nabla\times\mathbf{H})$, 
where $\beta$ characterizes the strength of the chirality. 
They can be brought to the same form. 
The transformations between the parameters of the two systems are given in Ref. 26.

\bibitem{CST}
CST MICROWAVE STUDIO (CST MWS) 
is a specialist tool for the 3D EM simulation of high frequency components, http://www.cst.com/Content/Products/MWS/Overview.aspx.

\bibitem{wavepropagation}
P. Marko\ifmmode \check{s}\else \v{s}\fi{} and C. M. Soukoulis, 
\textit{Wave Propagation: From Electrons to Photonic crystals and Left-Handed Materials} 
(Princeton University Press, Princeton, 2008). 

\bibitem{rkzhao}
R. Zhao, J. Zhou, Th. Koschny, E. N. Economou, and C. M. Soukoulis, ``Repulsive Casimir force in chiral metamaterials,''
Phys. Rev. Lett. \textbf{103}, 103602 (2009).

\bibitem{Pendry_mu}
J. B. Pendry, A. J. Holden, D. J. Robbins, and W. J. Stewart,`` Magnetism from conductors and enhanced nonlinear phenomena,''
IEEE Trans. Microwave Theory Tech. \textbf{47}, 2075 (1999).

\bibitem{condon}
E. U. Condon,``Theories of optical rotatory power,'' Rev. Mod. Phys. {\bf 9}, 432-457 (1937).

\bibitem{Landau}
L.D. Landau, E. M. Lifshitz, and L. P. Pitaevskii, 
\textit{Electrodynamics of Continuous Media} 
(2nd ed., Pergamon Press, Oxford, 1984), $\S$104, p.362-367.




\end{thebibliography}
\end{document}